\documentclass[12pt]{article}
\usepackage{epsfig}

\def\beq{\begin{equation}}
\def\eeq#1{\label{#1}\end{equation}}
\def\eeqn{\end{equation}}

\def\beqa{\begin{eqnarray}}
\def\eeqa#1{\label{#1}\end{eqnarray}}
\def\eeqan{\end{eqnarray}}

\let\bar=\overbar

\def\Dslash{\not{\hbox{\kern-4pt $D$}}}
\def\dslash{\not{\hbox{\kern-2pt $\del$}}}

\def\msb{{\bar{\ssstyle M \kern -1pt S}}}

\def\Title#1{\begin{center} {\Large {\bf #1} } \end{center}}

\begin{document}

\Title{Search for rare purely leptonic decays at LHCb}

\bigskip\bigskip

\begin{raggedright}  

{\it Flavio Archilli\index{Archilli, F.}\\
on behalf of the LHCb collaboration\\
Laboratori Nazionali di Frascati\\
Istituto Nazionale di Fisica Nucleare\\
I-00044 Frascati, ITALY}
\bigskip\bigskip
\end{raggedright}

\section{Introduction}

The aim of the LHCb experiment at the Large Hadron Collider at CERN is to perform precise tests 
of the Standard Model (SM) in the flavour sector, in order to disentangle  
possible New Physics (NP) effects. 
In particular the measurement of the branching fractions (BF) 
of the rare leptonic decays of the $B$, $D$ and $K$ mesons can give hints of the presence of 
NP particles at the tree and loop levels. 
The LHCb detector is well suited to study decays with muons in the 
final states: very efficient trigger allows to collect events containing one or two muons 
with very low transverse momenta; very good momentum resolution $\delta \mbox{p}/\mbox{p} = (0.4 \,\mbox{-}\, 0.6)\%$ 
reflects into excellent invariant mass resolution ($\sigma(M) \sim 25$~MeV for B two-body decays); the offline muon identification permits to have a 
good muon efficiency, $\epsilon \sim 90\%$,  for
a muon misidentification rate less than 1\% for $1 < p < 100 $~GeV/$c$.

In this paper the analyses of four purely leptonic rare decays with $\sim 1\,\mathrm{fb}^{-1}$ of $pp$ 
collisions collected by LHCb in 2011 at $\sqrt{s}=7$~TeV are presented. 

\section{$K_{S} \rightarrow \mu^+\mu^-$}

The rare decay $K_{S} \rightarrow \mu^+\mu^-$ are a very useful source of 
information on the short-distance structure of $\Delta S = 1$ Flavour Changing Neutral Current (FCNC) transitions.
This decay is suppressed in the SM and the prediction on its BF is 
$\mathcal{B}(K_{S} \rightarrow \mu^+\mu^-)_{SM} = (5.0\pm1.5)\times 10^{-12}$~\cite{ksmumu:theory1}~\cite{ksmumu:theory2}.
The current best limit on this decay, obtained in 1973, is equal to 
$\mathcal{B}(K_{S} \rightarrow \mu^+\mu^-) < 3.2\times 10^{-7}$ at 90\% of C.L.~\cite{ksmumu:prev}.
The contributions of NP to the BF, e.g. from light scalars, are allowed up to one order 
of magnitude above the SM expectation.

\begin{figure}[htb]
\begin{center}
\epsfig{file=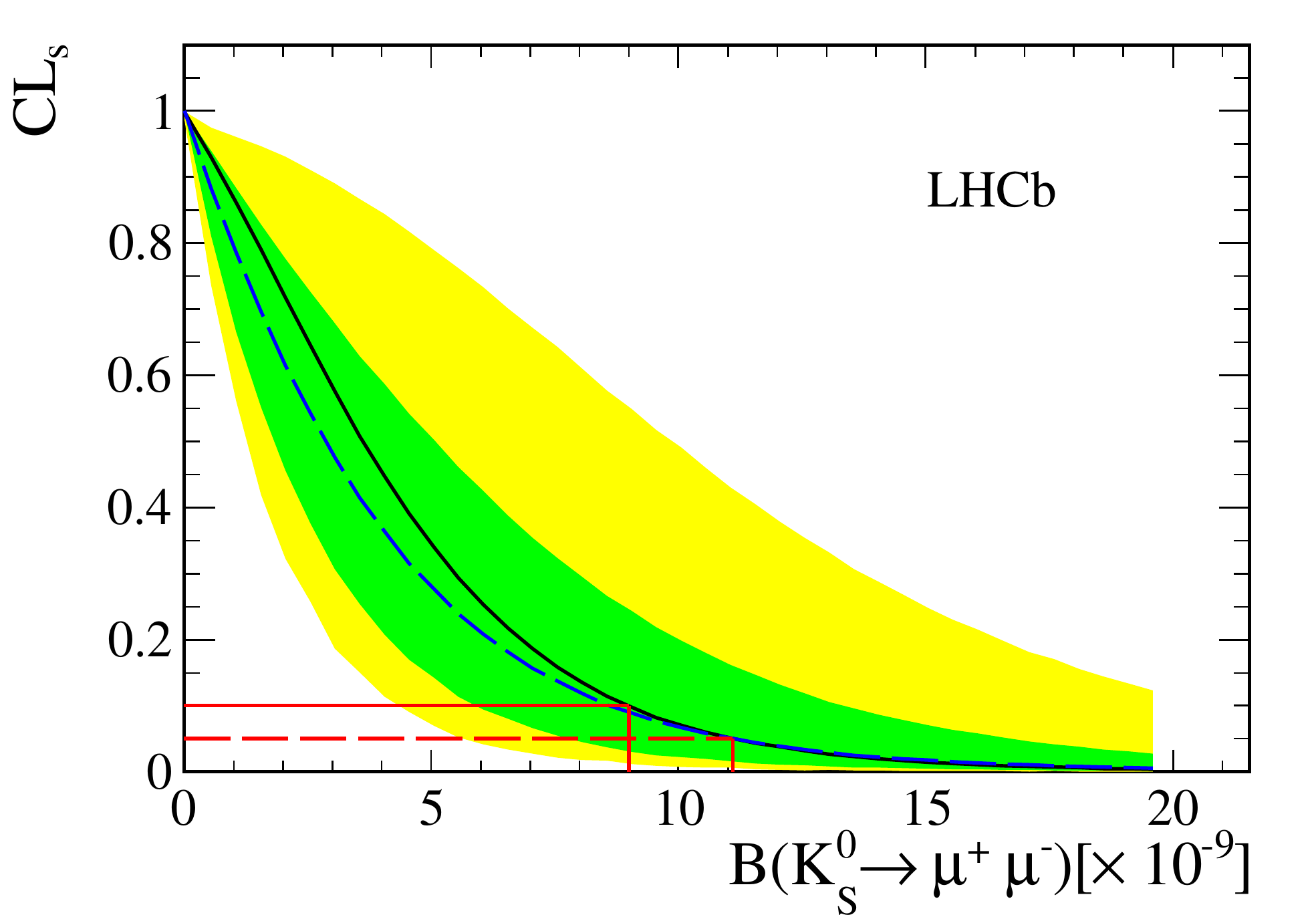,height=1.8in}
\caption{Expected $CL_s$ (dashed blue line) under the hypothesis to observe background-only. 
  The green (dark) band covers 68\% (1$\sigma$) of the $CL_s$ curves obtained in the background-only 
  pseudo-experiments, while the yellow (light) band covers 95\% (2$\sigma$). 
  The solid line corresponds to the observed $CL_s$.}
\label{ksmumu:fig1}
\end{center}
\end{figure}

A blind analysis is performed on $1\,\mathrm{fb}^{-1}$ of data collected during 2011 that contains $\sim 10^{13}$ $K_S$
inside the LHCb acceptance.
The main sources of background are due to combinatorial muons from
semileptonic decays and to $K_S\rightarrow \pi^+ \pi^-$ where both pions are misidentified as muons. The 
$K_L\rightarrow \mu^+ \mu^-$ component is negligible for this analysis.
The LHCb mass resolution is exploited to discriminate the $K_S\rightarrow \pi^+ \pi^-$ with both pions
misidentified as muons. Moreover, to increase the signal and background separation a multivariate classifier, 
a boosted decision tree (BDT), based on geometrical and kinematic informations is used.
The number of expected signal events, for a given branching ratio hypothesis, 
is evaluated by normalizing to the $K_S \rightarrow \pi^+\pi^-$ events. 
This normalization reduces the common systematic uncertainties between the two channels.
The modified frequentist method, $CL_s$, is used for the upper limit determination \cite{bsmumu:cls}.
The $CL_s$ curves for $\mathcal{B}(K_{S} \rightarrow \mu^+\mu^-)$ are shown in Fig.\ref{ksmumu:fig1}.
The observed upper limit is:
$
\mathcal{B}(K_{S} \rightarrow \mu^+\mu^-) < 9.0 \times 10^{-9} 
$
at 90\% C.L., with an improvement of a factor $\sim 30$ with respect the previous best limit.

\section{$B_{(s)}^0 \rightarrow \mu^+\mu^-$}

The search for $B_{(s)}^0 \rightarrow \mu^+\mu^-$ is one of the best
way for the LHCb experiment to constrain the parameters of NP models with an extended Higgs sector. 
These decays are highly suppressed in the SM because they are both FCNC and helicity suppressed.
The SM model predictions are \cite{bsmumu:Buras}:
$\mathcal{B}(B_{s}^0 \rightarrow \mu^+\mu^-)_{SM} = (3.2\pm0.2)\times 10^{-9}$ and 
$\mathcal{B}(B^0 \rightarrow \mu^+\mu^-)_{SM} = (1.0\pm0.1)\times 10^{-10}$. 
If we consider the Minimal Supersymmetric SM (MSSM) in the high $\tan \beta$ approximation, the enhancement
of $\mathcal{B}(B_{s}^0 \rightarrow \mu^+\mu^-)$ due to scalar and pseudoscalar sector is found to be proportional 
to $\tan^6 \beta$ \cite{bsmumu:tanb}.

\begin{figure}
\centering
 \begin{minipage}[b]{0.37\linewidth}
   \centering
   \includegraphics[width=\linewidth]{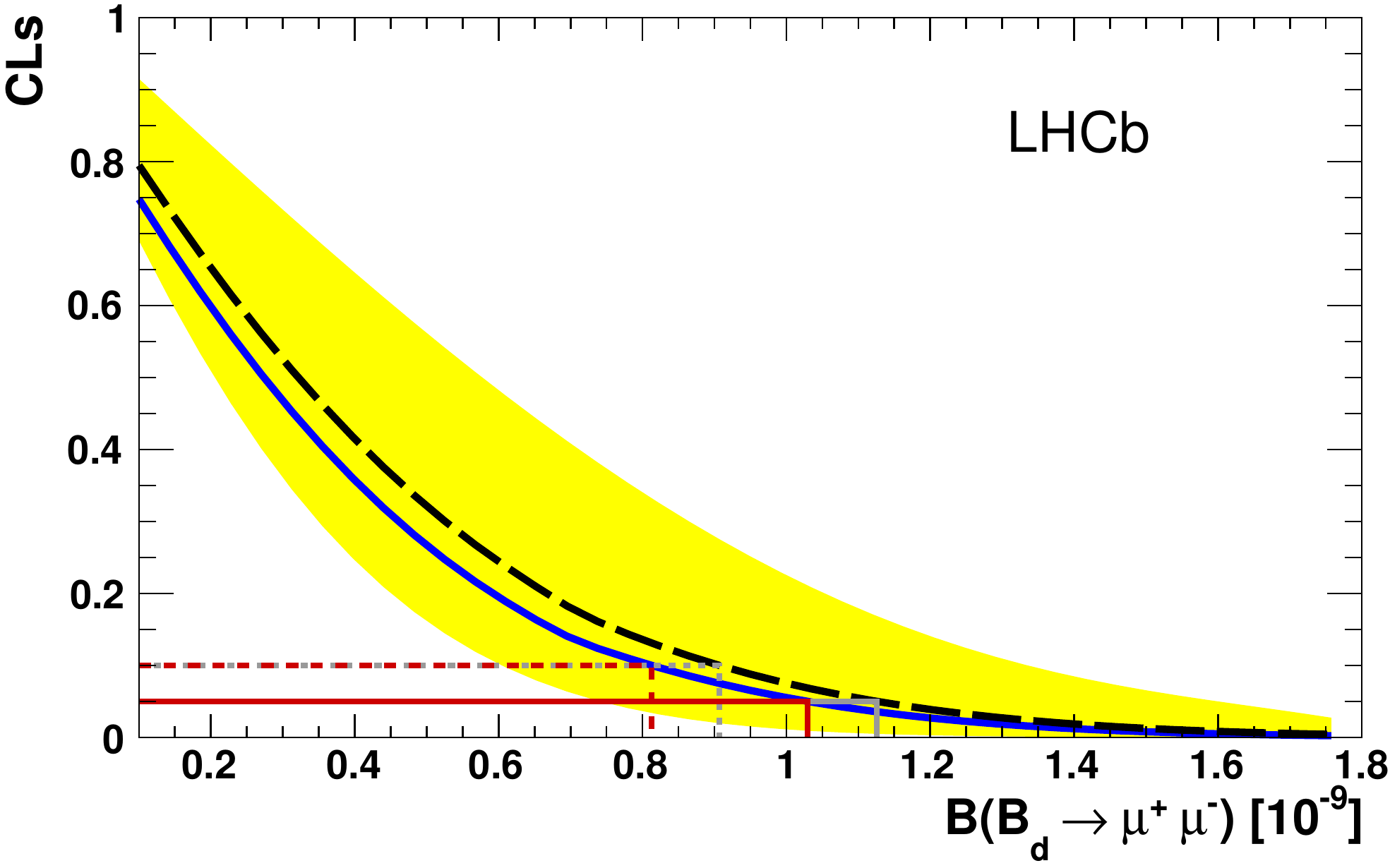}
 \end{minipage}
 \ \hspace{4.0mm} \
 \begin{minipage}[b]{0.37\linewidth}
  \centering
   \includegraphics[width=\linewidth]{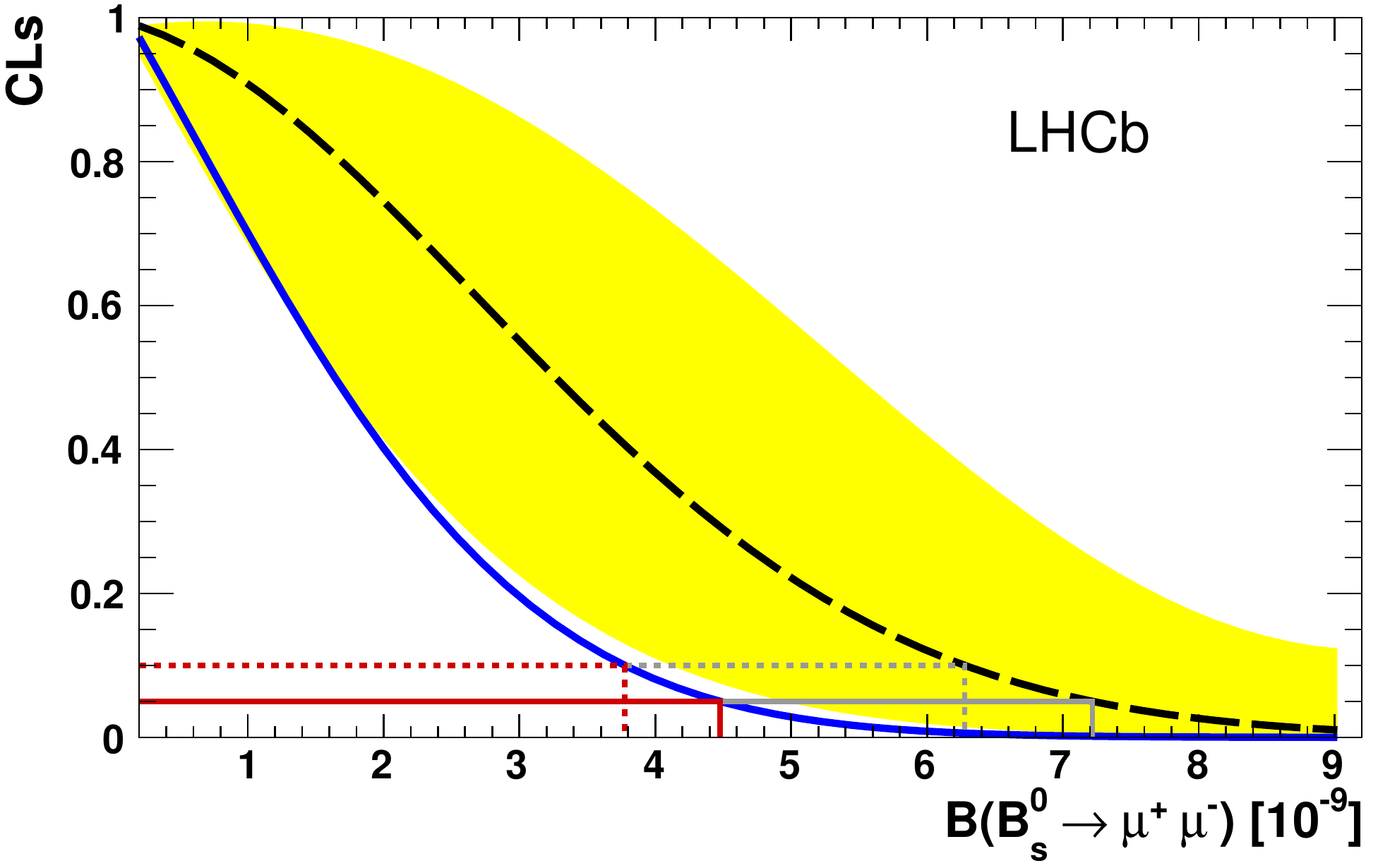}
 \end{minipage}
 \caption{Expected $CL_s$ (dashed black line) under the hypothesis to observe background-only (left) 
   for the $B^0 \rightarrow \mu^+\mu^-$ and $CL_s$ under background-plus-signal events according to the 
   SM rate (right) for $B_{s}^0 \rightarrow \mu^+\mu^-$, with yellow area covering the region of $\pm 1\sigma$ 
   of compatible observations; the observed $CL_s$ is given by the blue dotted line; the expected (observed) 
   upper limits at 90\% and 95\% C.L. are also shown as dashed and solid grey (red) lines.}
 \label{bsmumu:fig2}
\end{figure}

The results presented here are based on $\sim 1\,\mathrm{fb}^{-1}$ of data recorded by LHCb in 2011~\cite{bsmumu:prl}. 
The events are classified using a BDT, based on  kinematic and geometrical variables, and the dimuon invariant mass. 
In order to avoid any 
bias, the mass region $m_{\mu\mu} = [m(B^0)-60\,MeV/c^2,m(B^0_s)+60\,MeV/c^2]$ is blinded until the analysis is finalized. 
The number of expected signal events, for a given branching ratio hypothesis, 
is evaluated by normalizing to the decays $B^0 \rightarrow K\pi$, $B^+\rightarrow J/\psi K^+$ and 
$B^0_s \rightarrow J/\psi \phi$.
The compatibility of the observed events with background-only and background plus signal hypothesis is then computed. 
The modified frequentist method $CL_s$ is used for the upper limit determination \cite{bsmumu:cls}.
The $CL_s$ curves for $\mathcal{B}(B^0 \rightarrow \mu^+\mu^-)$ and $\mathcal{B}(B_{s}^0 \rightarrow \mu^+\mu^-)$ are shown in Fig.\ref{bsmumu:fig2}.

The upper limit obtained are:
$ 
\mathcal{B}(B_{s}^0 \rightarrow \mu^+\mu^-) < 4.5 \times 10^{-9}$, and $ 
\mathcal{B}(B^0 \rightarrow \mu^+\mu^-) < 1.03 \times 10^{-9}
$
both evaluated at 95\% C.L.. 
In order to compare the upper limit on $\mathcal{B}(B_{s}^0 \rightarrow \mu^+\mu^-)$ with the theoretical prediction, this value has to be multiplied by $0.911\pm 0.014$, which takes into account the effective lifetime of the $B_s$ meson~\cite{bsmumu:time}.

These two upper limits have been also recently combined \cite{bsmumu:comb} with the results obtained by 
CMS with 4.9$\,\mathrm{fb}^{-1}$ \cite{bsmumu:CMS} and ATLAS with 2.9$\,\mathrm{fb}^{-1}$ \cite{bsmumu:Atlas}. The combined upper
limits are: $\mathcal{B}(B_{s}^0 \rightarrow \mu^+\mu^-) < 4.2\times 10^{-9} $ and $\mathcal{B}(B^0 \rightarrow \mu^+\mu^-) < 8.1 \times 10^{-10}$ both at 95\% of C.L..
Despite the fact that the $\mathcal{B}(B_{s}^0 \rightarrow \mu^+\mu^-)$ limit is very close to the SM prediction, 
NP contributions are still possible.
For example, destructive interference between NP and the SM could decrease significantly the value of the $\mathcal{B}(B_{s}^0 \rightarrow \mu^+\mu^-)$.

\section{$D^0 \rightarrow \mu^+\mu^-$}

The $D^0 \rightarrow \mu^+\mu^-$ branching fraction is very suppressed in the SM. Its main contribution is 
due to two-photons intermediate state which sets the SM prediction bounds:
$10^{-13} < {\mathcal B} < 6\times10^{-11}$ at $90\%$ C.L. \cite{d0mumu:Burdman}. 
Many NP models predict enhancements for this BF, e.g. RPV-SUSY, which predicts $\mathcal{B}\sim 10^{-9}$ 
due to a tree level transition~\cite{d0mumu:Golowich}. The current best experimental limit is ${\mathcal B} < 1.4\times10^{-7}$ at $90\%$ C.L., 
from Belle \cite{d0mumu:Belle}.

An analysis has been performed on a data sample of 0.9$\,\mathrm{fb}^{-1}$ and selecting the decay chain
$D^{*\pm}\rightarrow D^0(\rightarrow \mu^+\mu^-)\pi^\pm$.
The selection of the events is based on geometric and kinematic properties of $D^0$ and its daughter particles. 
After this selection the background is mainly composed of combinatorial muons from
semileptonic decays of $b$ and $c$ hadrons and the peaking background of $D^{*\pm}\rightarrow D^0(\rightarrow h^+h^-)\pi^\pm$ events, with double
hadron misidentification. The combinatorial background is reduced making use of a 
multivariate discriminant based on geometrical and kinematical information.
The signal events are normalized to the $D^{*\pm}\rightarrow D^0(\rightarrow \pi^+\pi^-)\pi^\pm$
channel. Since the same selection is applied to signal and normalization channel, 
this procedure permits to reduce common systematic uncertainties in both channels.
The event yield is determined from a 2D fit on the dimuon invariant mass and the difference between the $D^{*+}$ mass and $D^0$ mass.
A preliminary result is given:
$\mathcal{B}(D^0 \rightarrow \mu^+\mu^-) < 1.3 \times 10^{−8}\mbox{ at 95\% C.L.}$ \cite{d0mumu:LHCb}.

\section{$B_{(s)}^0 \rightarrow \mu^+\mu^-\mu^+\mu^-$}

The $B_{(s)}^0 \rightarrow \mu^+\mu^-\mu^+\mu^-$ is a FCNC process and takes its largest contribution 
from the resonant decay $B_s \rightarrow J/\psi \phi$ in which both mesons decay into two muons. 
In the SM a non-resonant process can also occur through the exchange of a virtual photon with a 
$BF\sim 10^{-10}$~\cite{bsmumumumu:Melikhov}. However, NP effects might enhance the 
$ \mathcal{B}(B_{(s)}^0 \rightarrow \mu^+\mu^-\mu^+\mu^-)$ by the exchange of new particles
at the tree level.

A cut-based selection has been designed on the 1$\,\mathrm{fb}^{-1}$ data sample collected during the year 2011. 
The selection algorithm is tuned using the resonant decay mode and optimized keeping blind the signal region.
The combinatorial background was estimated from the mass sidebands.
The selection criteria for signal and control channels are based on particle identification, separation 
between the B vertex and the primary vertex, the quality of the B decay vertex. A veto on $\phi$ and $B^+$ masses 
is applied in order to exclude $B^+ \rightarrow J/\psi K^+$ and $B \rightarrow K^* \phi$.
All the non-resonant peaking background yields in the signal region are found to be negligible.

The signal BF is measured by normalizing to 
$B_s^0 \rightarrow J/\psi(\rightarrow \mu^+\mu^-)K^{*0}(\rightarrow K^+\pi^-)$ decays selected 
with the same criteria. Systematic uncertainties are evaluated by comparing Monte Carlo simulation with data.

After unblinding, one event is observed in the $B_d$ signal window and no events are
observed in the $B_s$ window.
The number of observed events is consistent with the expected background yields.
The $CL_s$ method has been used to evaluate the upper limits on BF~\cite{bsmumumumu:LHCb}:
${\mathcal B} (B_{(s)}^0 \rightarrow \mu^+\mu^-\mu^+\mu^-)< 1.3\,\times \, 10^{-8}$ and
${\mathcal B} (B^0 \rightarrow \mu^+\mu^-\mu^+\mu^-)< 5.4\,\times \, 10^{-9}$ both  at 95\% C.L.

\section{Conclusion}

We presented some of the purely leptonic rare decays analyses based on $1\,\mathrm{fb}^{-1}$  collected by LHCb during the 2011 run.
A brand new result on the upper limit on $\mathcal{B}(K_S\rightarrow \mu^+\mu^-)$ is presented. This result improves
the previous best limit by a factor $\sim 30$.
The results on the upper limits on $\mathcal{B}(B^0\rightarrow \mu^+\mu^-)$ and $\mathcal{B}(B_{s}^0\rightarrow \mu^+\mu^-)$ 
put severe constraints on NP and further improvements are foreseen.
The combination of the upper limits obtained by ATLAS, CMS and LHCb has been also reported. This 
combination improves on the limits obtained by the individual experiments and represents the best existing limits on these decays.
The study on the upper limit on $\mathcal{B}(D^0\rightarrow \mu^+\mu^-)$ shows an improvement on the previous best limit by an order of
magnitude.
A search for the decays $B^0_{s}\rightarrow\mu^+\mu^-\mu^+\mu^-$ and $B^0\rightarrow\mu^+\mu^-\mu^+\mu^-$ is also discussed and the upper limits on their branching fractions are reported. 
Many improvements are foreseen in the near future. With the 2012 run LHCb expects to more than double the data
set and will put even more stringent constraints on the phase space of many NP models.

\end{document}